# Towards Greener & Safer Mines

*Dhruv Srivastava[a] and Dr. Priya Ranjan[b]*

*a. B .Tech. III Yr, Department of Electronics Engineering, Indian School of Mines University, Dhanbad – 826004, India, E-mail: dhruv15125@ismu.ac.in*

*b. Assistant Professor, Department of Electrical Engineering, Indian Institute of Technology, Kanpur – 208016, India, E-mail: pranjan@gmail.com*

**Abstract**

Miniaturised sensors and networking are technical proven concepts. Both the technologies are proven and various components e.g., sensors, controls, etc. are commercially available. Technology scene in above areas presents enormous possibilities for developing innovative applications for real life situations. Mining operations in many countries have lot of scope for improving environmental and safety measures. Efforts have been made to develop a system to efficiently monitor a particular environment by deploying a wireless sensor network using commercially available components. Wireless Sensor Network has been integrated with telecom network through a gateway using a suitable topology which can be selected at the application layer. The developed system demonstrates a way to connect wireless sensor network to external network which enables the distant administrator to access real time data and act expediently from long-distance to improve the environmental situation or prevent a disaster. Potentially, it can be used to avoid the awful situations leading to terrible environment in underground mines.

### 1. Introduction

The impact of internet, cell phones, and other communication modes has been enormous. Tremendous growth in both sensor network technology [1-3] and other applications are driving the need for new techniques and tools for analyzing and visualizing sensor network data streams. Convergence of wireless sensor networks with telecommunication is not far away and presents enormous possibilities for developing innovative applications for real life situations which may also bring commercial services like security updates of one's home on your cell phone [4, 5]. One of the most significant applications of wireless sensor networks is environment monitoring in harsh and inaccessible places like mines, nuclear reactors, etc [6].

Mines have a very hazardous environment not only from environmental angle but also from human security point of view. Hundreds of people die every year in the mines due to the lack of safety measures [7]. Unfortunately, the uptake of technology is slow in the mining sector. Astonishingly, things have been so terrible from the past few decades.

One of the major issues in the mines especially the underground mines is the concentration of oxygen, and other gases like carbon monoxide, nitrogen, and methane [8, 9]. The inappropriate concentration of these gases is unacceptable from ergonomic and environmental angle and even lead to explosion which creates greenhouse gases. If these parameters can be remotely monitored from long distance by the decision makers then required actions can be taken to prevent the environment damage. In general, the person responsible for managing the workers and labors who work in the mines are far away from the actual mines. These managers are more interested in keeping the mining processes going on to improve the productivity irrespective of its impact on the strategic concerns like environment. By using our proposed way harmful situations in the mines can be known to the senior enlightened executives or who are trained professionals can take call.

Integrating a wireless sensor network having gas sensors with a telecom network will enable remote monitoring of the nature and concentration of these gases and risks can be mitigated through appropriate executive orders in real time. However, only temperature and light sensors were used for demonstrating the concept. Thus by deploying a Wireless sensor network integrated with IP network, the conditions inside the active mines can be monitored from a remote computer.

2. **Topologies for the wireless sensor network**

The network topology to be selected for the wireless sensor network depends on the basis of the area to be covered, data redundancy or energy optimization issues. Like, ring topology, star topology, mesh topology etc. Every network topology has its own advantages and disadvantages. Hybrid topologies can also be implemented on the wireless sensor network. Good examples of such topologies are, star on ring topology preferred when distance coverage needs to be increased and star on star topology which is preferred to avoid data redundancy and achieve faster communication. We have implemented a tree topology on the wireless sensor network because data is collected and passed on very efficiently without much data redundancy. Also, it is suitable for enhancing the network coverage area by adding up new nodes to the network. *Figure1* shows a data flow diagram in tree topology.

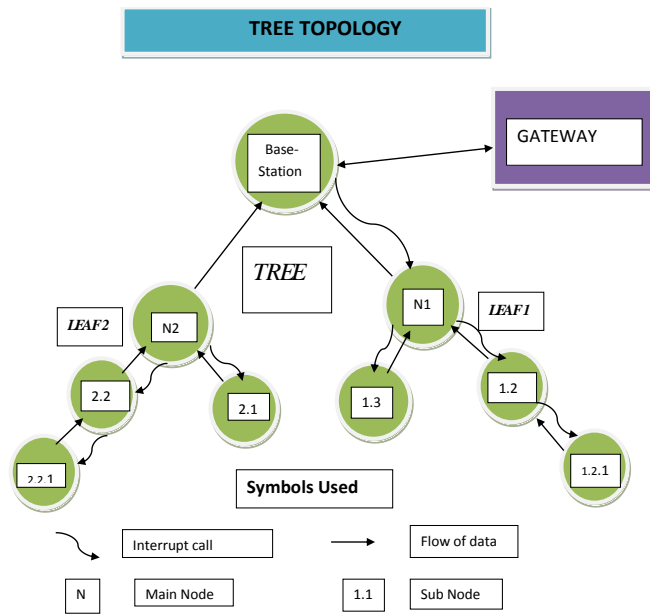

*Figure1: Data flow diagram for a wireless sensor network in tree topology*

### 3. Implementation of the proposed method

The wireless sensor network needs to be setup in a suitable network topology depending upon the application it has to be used for. We have implemented a tree network topology for our demonstration of our work. The network has cluster heads which collects data from every leaf node and then these cluster heads sends the whole data to the base station which is connected to an external network through a gateway. However, only temperature and light sensors were used for demonstrating the concept .Sensors, wireless technologies and computer networking tools have been used. It has been demonstrated, how by our concept data from the wireless sensor network can be accessed expediently from long-distances which have been illustrated in *figure2.*

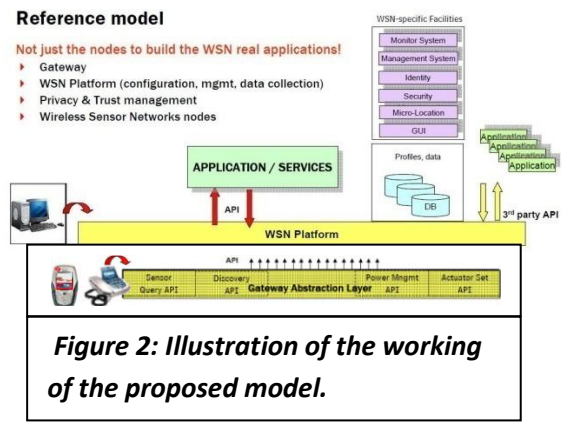

*Figure 2: Illustration of the working of the proposed model.*

As mentioned in previous section, we have used a tree topology for the wireless sensor network. The network is having 2 main leaves (cluster heads) & their 2 leaflets (sub-nodes) each. In total we monitored the values of temperature & light intensity of 6 nodes in total. **Figure 3** shows a data flow chart of what exactly have we implemented*.* We also plotted graphs for both the readings i.e. temperature & light intensity coming from every node. The graphs being plotted at the time of the experiment can be seen in **Figures 4-9.** The readings of temperature & light intensity coming from every node were taken from the base station and written onto a file (.txt). The file was being updated at regular interval of time and was sent to different clients on their request, respectively. **Table 1** shows the software and hardware used for performing the experiment; their detailed description has been reported

[10] and can be found in their respective datasheets also.

*Terminology used:*

i. *Interrupt call-* When a node sends a message to another node asking to send the data it has.
ii. *Flow of data-* Sending information from node to another i.e. wireless communication between the nodes.
iii. *Topology-* The way/structure in which the network has been set up.
iv. *Leaf-* Used as a synonym for a node in tree topology.
v. *Leaflet-* Used as a synonym for a sub- node in tree topology.

*Assumption:* Wireless range of communication of every node is nearly same.

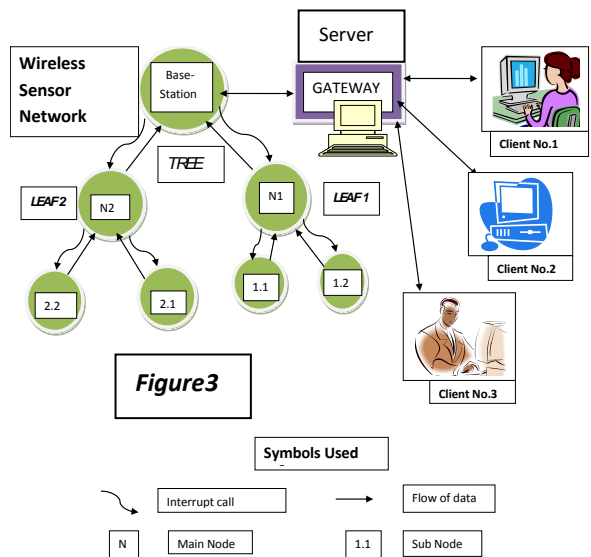

*Figure 3: The experimental set up: A flow chart of what exactly have we implemented.*

## 4. Results and Discussions

We successfully implemented a multi-hop network of sensor nodes as shown in *figure 3* which are capable of monitoring temperature, light intensity and other ambient data. The readings of temperature & light intensity coming from every node placed at various location of the laboratory were the experiment was performed. The readings were taken from the base station as it has data coming from every node. Then the readings were plotted on graphs, which are shown in *figures 4-9.*

The base station of the deployed network is responsible for aggregation of data from all nodes and sending it to the main server via a gateway. The server can provide service to multiple clients. A customized application layer has been built on the client machines to facilitate efficient data monitoring and cluster selection from a deployed wireless sensor network. This server can provide facilities to the telecom servers and application layer interface can be built on PDAs, mobile phones, etc, thus increasing the number of clients and ease of information access. This will lead to emergence of a new generation of telecommunication value added services and provide facilities for building automation, health monitoring, advanced metering application etc.

The same setup can be used to enhance mine safety as already mentioned. The wireless sensor network must be planted in a mine. The nodes of the network must be placed at various locations of the mine. Each node should be planted at different locations of the mine covering the whole area. Each node needs to be fabricated with temperature sensors like TMP-275 and various gas sensors like TGS-2611, TGS-2442, and TGS-2600 which can measure the parameters like temperature, concentration of oxygen, carbon monoxide nitrogen, and methane.

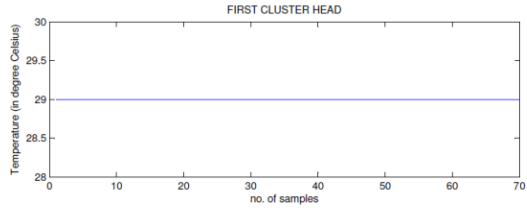
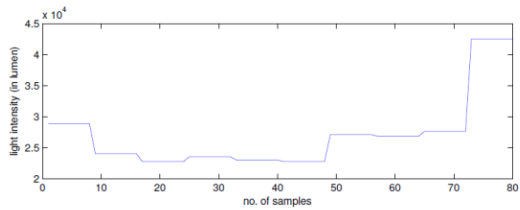

*Figure 4: Readings of node N1- second cluster head*

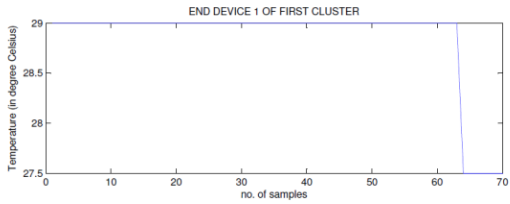
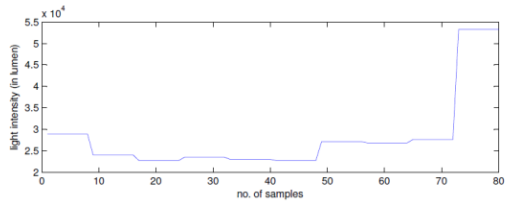

*Figure 5: Readings of node 1.1- endevice1 of first cluster head*

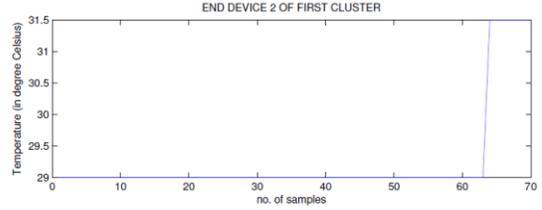
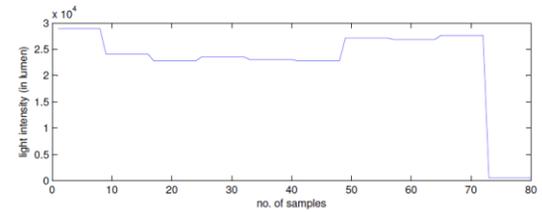

*Figure 6: Readings of node 1.2- endevice2 of first cluster head*

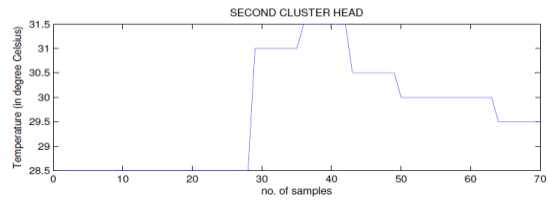
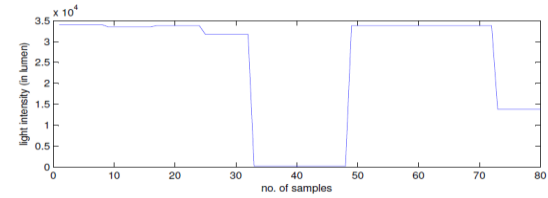

*Figure 7: Readings of node N2- second cluster head*

*Figure 8: Readings of node 2.1- endevice1 of second cluster head*

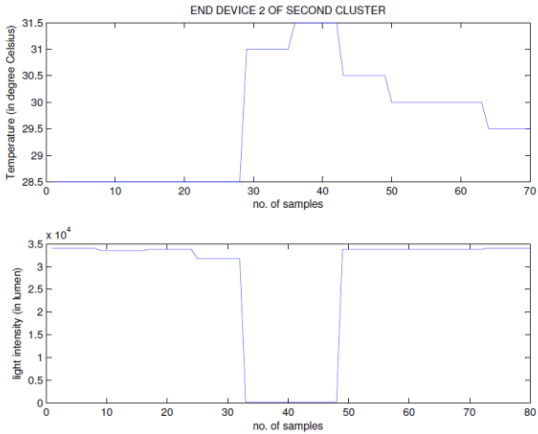

*Figure 9: Readings of node 2.2-endevice2 of second cluster head*

There is not much of variation in the temperature readings because all nodes were placed in the same room. There are a very few points where the temperature or the light intensity readings have suddenly dropped down to null. This probably happened because of the communication breakage between the nodes i.e. between the leaf and leaflets due to unknown reason. We were not able to avoid it. But important part is that the readings can be monitored and can be seen on the graphs.

## 5. Conclusions

We have successfully implemented a multi-hop network of sensor nodes which are capable of monitoring temperature & light intensity. The base station for this network is responsible for aggregation of data from all nodes and sending it to the main server via a gateway. The server then provides service to multiple clients.

This small feasibility experiment makes ways in the following directions:

a. Mining safety applications
b. Environment monitoring
c. Bio-hazard monitoring
d. Police and security applications
e. Domestic applications

We already have reached a level where we have shown the effectiveness of wireless sensor network with available set of nodes. In addition to it, most of the software made and used is open source further modifications can be made as per the requirements. We strongly believe that this work can be taken to a much bigger platform which can bring drastic changes in the present situation of mines safety and environment.

## Acknowledgements

Vividh Mishra, Senior Project Associate, *Department of Electrical Engineering, Indian Institute of Technology, Kanpur – 208016, India*.

Table-1: Software and Hardware used with their brief description

| | | |
|---|---|---|
| AVR Studio 4 | Software | For programming & debugging the AVR microcontroller. |
| X-CTU | Software | Allows the ability to program the radios' firmware settings via a graphical user interface.(Windows based) |
| MATLAB (optional) | Software | For serial port interfacing and plotting readings on graphs. |
| Indriya CS-03A14 Kit | Hardware | Wireless sensor module having options to plug in different combinations of sensors[1] |
| AVR Atmega 128L microcontroller | Hardware | 7.3728 MHz, 128KB flash, 4KB RAM processing unit |
| Temperature Sensor (TMP-275) | Hardware | $0.5^0C$ accuracy digital response |
| Ambient Light Sensor (APDS-9300) | Hardware | approximate human eye 16bit I2C compatible response |
| XBee Wireless Radio (2.4 Ghz) | Hardware | IEEE802.15.4 compliant radio with 30m-100m range |
| USB-UART Interface(FTDI chip) | Hardware | For communication of hardware with a computer through serial communication |